\documentstyle[aps,prl,preprint,epsfig]{revtex}
\begin{document}
\draft
\noindent {\Large \sffamily \bf 
Superconducting energy gaps, low temperature specific heat, and 
quasiparticle spectra of MgB$_2$
}\\
\\
\noindent {\sffamily Hyoung Joon Choi,$^*$ David Roundy,$^{*\dagger}$ Hong Sun,$^{*}$
Marvin L. Cohen,$^{*\dagger}$ \& Steven G. Louie$^{*\dagger}$}\\
\noindent {\em $^*$Department of Physics, University of California at Berkeley,
Berkeley, CA 94720, USA.\\
$^\dagger$Materials Sciences Division, 
Lawrence Berkeley National Laboratory, Berkeley, CA 94720, USA.
}

{\bf
Magnesium diboride is remarkable not only for its
unusually high transition temperature of 39K
for a $sp$-bonded metal\cite{nagamatsu01} 
and its reduced isotope-effect exponent\cite{budko01,hinks01},
but also for its unusual superconducting gap 
behavior\cite{karapetrov01,sharoni01,rubio01,schmidt01,takahashi01,szabo01,giubileo01,laube01,wang01,bouquet01,yang01,chen01,tsuda01}.
Some earlier experiments seem consistent with conventional single-gap 
superconductivity\cite{karapetrov01,sharoni01,rubio01,schmidt01,takahashi01},
but recent measurements indicate the existence of more than one 
gap\cite{szabo01,giubileo01,laube01,wang01,bouquet01,yang01,chen01,tsuda01}.
For example, specific heat measurements show an anomalous large bump
at low temperature, inconsistent with single-gap BCS theory.
Moreover, there is a significant variation in the measured values for the gaps.
Here, we report first-principles calculations of 
the $k$- and $T$-dependent superconducting gap $\Delta(\vec{k},T)$
in MgB$_2$ and its manifestation in various measured quantities.
Because its Fermi surface has disconnected sheets 
with very different electron-phonon coupling strengths,
our calculations show that near $T=0$, the values of $\Delta(\vec{k})$ 
cluster into two groups: large values ($\sim$ 6.5 to 7.5 meV) 
on the strongly coupled sheets and
small values ($\sim$ 1 to 3 meV) on the weakly coupled sheets.
The calculated gap,
quasiparticle density of states, and specific heat and  their 
temperature dependences are in agreement
with the recent measurements which support that MgB$_2$ is a multiple
gap superconductor. In fact, theory predicts four prominent values
for the gap at low $T$.}

Most theoretical work on MgB$_2$  has been focused 
on the electronic structure, the phonon structure,
and the electron-phonon 
interaction\cite{kortus01,an01,bohnen01,yildirim01,liu01,kong01,choi01}.
It is shown that the electron-phonon interaction in MgB$_2$ 
varies strongly on the Fermi surface\cite{liu01,kong01,choi01}.
Also, recent two-band model\cite{liu01} and
first-principles\cite{choi01} studies of the transition temperature 
support a multigap scenario\cite{suhl59}.
However, no detailed quantitative calculations
have been presented for the temperature and $k$-space dependence of the
superconducting gap.
The calculational framework\cite{choi01} used here  
is based on the strong coupling formalism
of superconductivity established by Eliashberg\cite{eliashberg60,allen82}.
The Eliashberg formalism is a more general case of the original 
formulation of the BCS theory for 
phonon-mediated pairing, taking into account strong electron-phonon
coupling.
When applying the strong coupling theory to MgB$_2$, 
we include the momentum dependency of the anisotropic
Eliashberg equation and also
take into account the anharmonic effect on phonon frequencies.
Material-dependent parameters in the equation are obtained 
by {\em ab initio} pseudopotential density functional 
calculations.
This approach has already been shown to describe
the superconducting transition
of MgB$_2$ very accurately; the obtained specific heat mass enhancement factor
$\lambda = 0.61$, the transition temperature $T_c = 39K$, and 
isotope-effect exponent $\alpha_B = 0.31$ are all in excellent 
agreement with experiments\cite{choi01}.
Here, we compute the gap function $\Delta(\vec{k},\omega)$ 
and the renormalization factor $Z(\vec{k},\omega)$ 
from the fully anisotropic Eliashberg equation
at various temperatures. This allows us to address
 all the superconducting properties
except those related to the presence of an applied magnetic field.

Figure 1 shows the calculated superconducting gap $\Delta(\vec{k})$ 
on the Fermi surface at 4 K. 
The Fermi surface of MgB$_2$ consists of four sheets:
two-dimensional (2D) light hole and heavy hole sheets forming 
coaxial cylinders along  $\Gamma$ to $A$, 
a three-dimensional (3D) hole sheet connecting regions near $K$ and $M$,
and a 3D electron sheet connecting regions near $H$ and $L$.
The 2D light and heavy hole sheets are derived from $\sigma$-antibonding 
states of boron $p_{x,y}$ orbitals, while 
the 3D hole and electron sheets are derived from $\pi$-bonding 
and antibonding states of boron $p_z$ orbitals, respectively.
The calculated density of states at the Fermi energy is 
0.115~states/eV$\cdot$atom$\cdot$spin; 44 \% of which
comes from the 2D $\sigma$ cylindrical
sheets and the rest comes from the 3D $\pi$ sheets.
The superconducting gap is nonzero everywhere on
the Fermi surface. Thus, the symmetry of the superconducting gap is s-wave,
but the size of the gap changes on the Fermi surface.
The gap values cluster into two groups.
The largest gap is on
the 2D light hole $\sigma$ cylindrical sheet (shown red in Fig.~1),
where the average gap is  7.2 meV with variations of less than 0.1 meV.
On the 2D heavy hole $\sigma$ cylindrical sheet (shown orange in Fig.~1), the superconducting gap ranges
from 6.4 meV to 6.8 meV, having an average of 6.6 meV, with
maximum value near $\Gamma$ and minimum value near A. 
The average of the gap values 
on the two 2D $\sigma$ cylindrical sheets is 6.8 meV.
The superconducting gap is significantly smaller and 
more spread out on the 3D $\pi$ sheets (shown green and blue
in Fig.~1), ranging from 1.2 meV
to 3.7 meV.
Out of the two 3D $\pi$ sheets, 
the hole sheet shows a slightly larger gap with a peak at 2.1 meV and
the electron sheet shows a somewhat smaller gap with a peak at 1.5 meV.
The average of the gap values on the two 3D $\pi$ sheets is 1.8 meV.
This very large variation in the gap value originates from 
the differences in electron-phonon coupling strengths on the different
sheets of the Fermi surface. Our {\em ab initio} calculations\cite{choi01}
 demonstrate that the electron-phonon
coupling strength is 4 to 5 times stronger on the cylindrical $\sigma$ sheets
(exceeding 2.5 for some states) than on the $\pi$ sheets.
Our result is consistent with the recent experiments reporting two 
gaps\cite{szabo01,giubileo01,laube01,wang01,bouquet01,yang01,chen01,tsuda01}
because
the superconducting gap in MgB$_2$  can be grossly viewed 
as consisting of a large gap of $\sim$ 6.8 meV 
on the strongly coupled 2D $\sigma$-cylindrical sheets  and
a small gap of $\sim$ 1.8 meV on the weakly coupled 3D $\pi$ sheets.
However, this simplication is limited.
As seen in Fig.~1, the large and small gaps 
both have substantial variation 
in value: 6.4 to 7.2 meV on the 2D $\sigma$ sheets 
and 1.2 to 3.7 meV on the 3D $\pi$ sheets.
The experimental interpreted gaps have a range of 1.5 to 3.5 meV for the 
small gap and a range of 5.5 to 8 meV for the large gap\cite{szabo01,giubileo01,laube01,wang01,bouquet01,yang01,chen01,tsuda01,buzea01}.

Figure~2 shows the temperature dependence of the superconducting energy gap
on the Fermi surface from 4K to 38K which may be probed in tunneling,
optical, and specific heat measurements.
The vertical blue curves in Fig.~2 present the distribution 
of the superconducting gap values at various temperatures,
and the red lines are curves of the form
$\Delta(T) = \Delta(0) \sqrt{1-(T/T_c)^p}$
fitted separately to the averaged values of the calculated 
gap on the 2D $\sigma$ 
sheets and those on the 3D $\pi$ sheets.
Our calculated results give
$\Delta(0) = 6.8$ meV ($2\Delta(0)/k_BT_c = 4.0$)
for the averaged gap at $T=0$ on the 2D $\sigma$ sheets, and 
$\Delta(0) = 1.8$ meV ($2\Delta(0)/k_BT_c = 1.06$)
for that on the 3D $\pi$ sheets.
The theory shows that the superconducting gap opens at $T_c$ 
throughout the Fermi surface, and that the gap 
widens much faster on the 2D $\sigma$ sheets than on the 3D $\pi$ sheets
at $T~ {}^<_{\sim}~ T_c$. 

From the calculated gap function $\Delta(\vec{k},\omega)$,
the quasiparticle density of states is given by
\begin{equation}
N(\omega)/N(0) 
= \mbox{Re}\left\langle\frac{\omega+i\Gamma}
{\sqrt{(\omega+i\Gamma)^2-\Delta(\vec{k},\omega)^2}}\right\rangle~,
\end{equation}
where $\langle\cdots\rangle$ indicates an average over a surface
of constant $\omega$.
Figure 3 depicts the theoretical quasiparticle density of states calculated
with an assumed finite lifetime $\Gamma$ of 0.1 meV.
The quasiparticle density of states at 4K shows three discernable peaks.
One peak is at 2.2 meV, which is
slightly greater than the dominant size of the gap (2.1 meV)
on the 3D $\pi$ hole sheet. The other two peaks are at 6.8 and 7.2 meV, which
are the maximal sizes of the superconducting gap on the 2D heavy 
and light hole $\pi$ sheets, respectively. 
The quasiparticle density of states can be deduced 
from tunneling experiments and various spectroscopic 
measurements\cite{szabo01,giubileo01,laube01,chen01,tsuda01},
but a direct quantitative
comparison requires knowledge of various physical parameters
involved in a specific experiment.

A fundamental measurement which can provide a probe of the
superconducting quasiparticle spectrum is the temperature-dependent
specific heat.
Figure 4 shows the calculated $T$-dependent specific heat of MgB$_2$
through the superconducting transition.
For the normal-state specific heat $C_N = \gamma T$,
we obtain $\gamma$ = 2.62 mJ/mol$\cdot$K$^2$\cite{choi01},
which agrees well with experimental values 2.6 mJ/mol$\cdot$K$^2$\cite{bouquet01}
and 2.7 mJ/mol$\cdot$K$^2$\cite{wang01,yang01}.
For the superconducting-state specific heat $C_S$,
we first calculate the
free energy difference ($F_S-F_N$) of the superconducting
and normal states\cite{bardeen64} and then
obtain the specific heat difference by
\begin{equation}
C_S-C_N = -T\frac{d^2}{dT^2}(F_S-F_N)~.
\end{equation}
The measured low $T$  specific heat\cite{wang01,bouquet01,yang01}
shows substantial magnitude and a large bump at about 10K.
This anomalous behavior is inconsistent with a 1-gap BCS model.
However, our results are in excellent agreement with experiment.
The origin of the bump in our calculated curve is
the existence of low energy excitations above the small 
gap on the weakly coupled 3D $\pi$ sheets of the Fermi surface.
The overall shape and magnitude 
of the calculated specific heat curve agrees very well
with experiments, especially below 30K, and shows
a jump of $\Delta C/\gamma T_c = 1.0$ at $T_c$ which
is within the range of the measured values.

This work was supported by National Science Foundation Grant No. DMR00-87088,
and by the Director, Office of Science, Office of Basic Energy Sciences
of the U. S. Department of Energy under Contract DE-AC03-76F0098.
Computational resources
have been provided by the National Science Foundation at the National Center
for Supercomputing Applications and by the National Energy Research Scientific
Computing Center. 
Authors also acknowledge financial support from the Miller Institute
(H.J.C.) and from the Berkeley Scholar Program funded by the Tang Family 
Foundation (H.S.).

Correspondence should be addressed to S.G.L. (e-mail: sglouie@uclink.berkeley.edu).

\newpage

\begin{figure}
\centering
\mbox{\epsfig{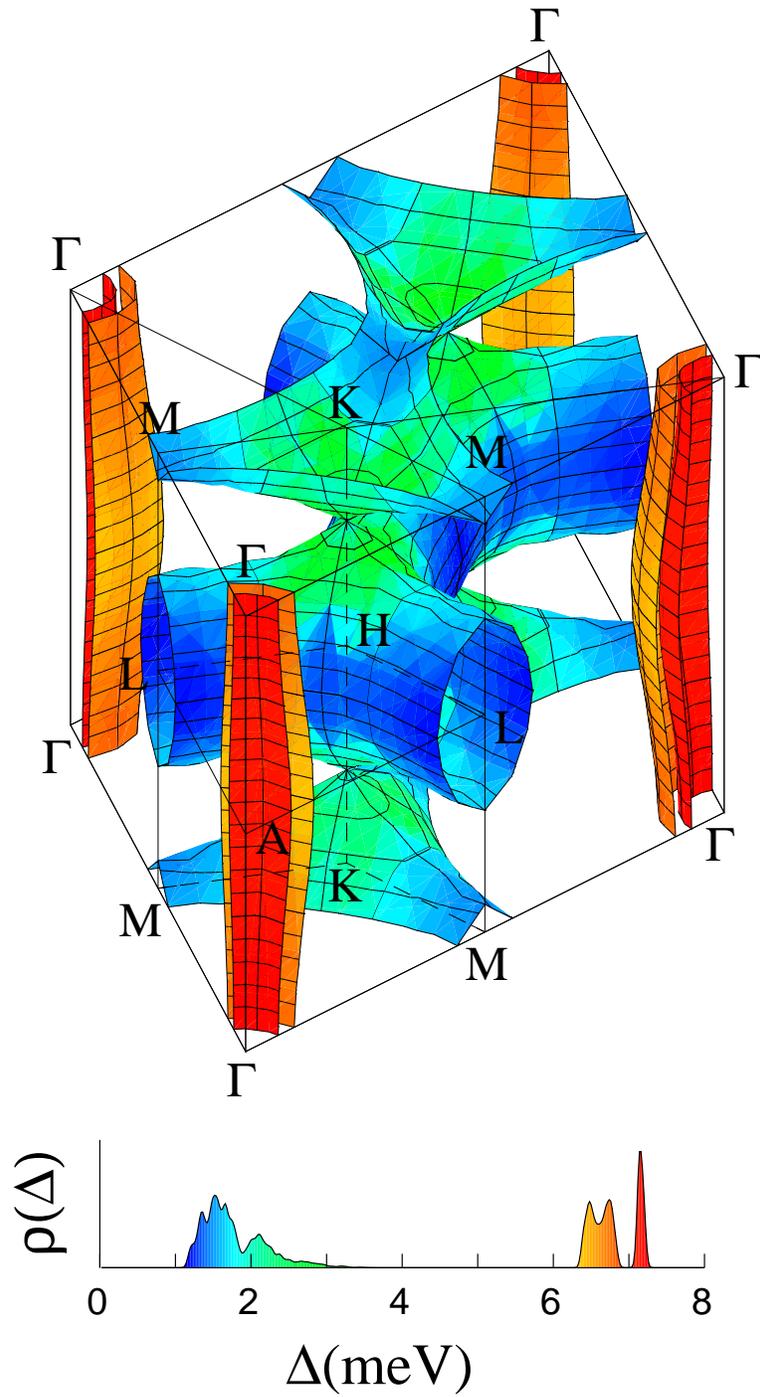}}
\caption{(color) The superconducting energy gap on the Fermi surface at 4 K.
Data is given in a color scale (top panel). The bottom panel depicts
the distribution of gap values and the color scale used.}
\end{figure}

\newpage
\begin{figure}
\centering
\mbox{\epsfig{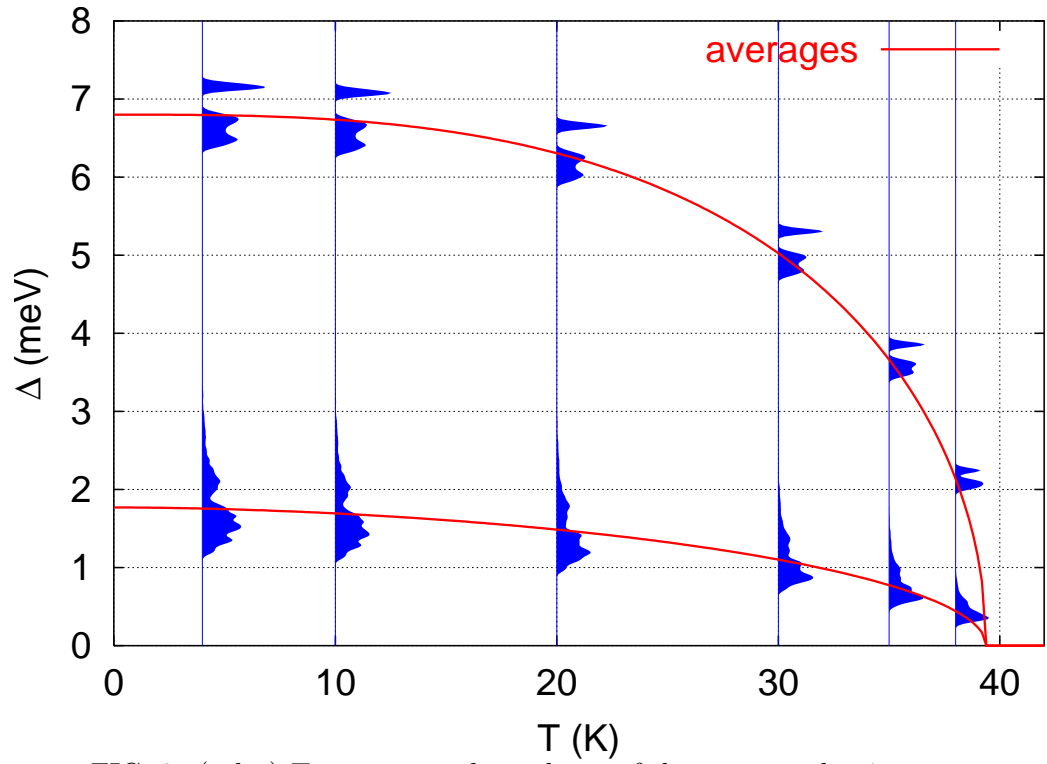}}
\caption{(color) Temperature dependence of the superconducting gap.}
\end{figure}

\newpage
\begin{figure}
\centering
\mbox{\epsfig{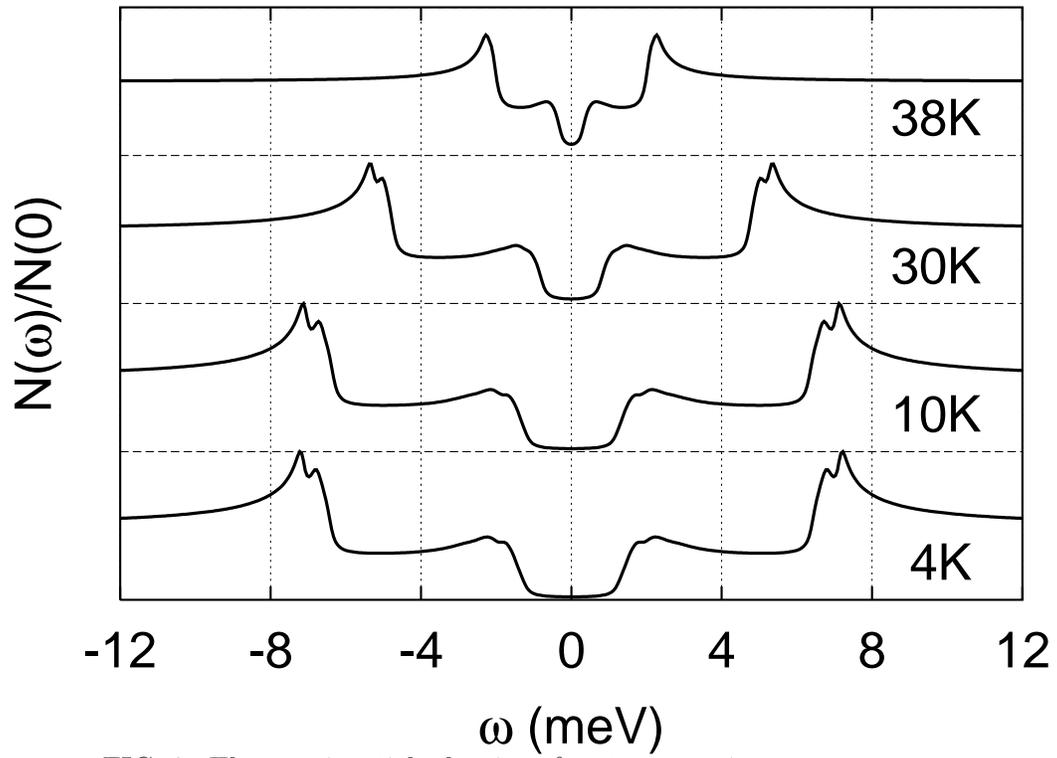}}
\caption{The quasiparticle density of states at various temperatures.}
\end{figure}

\newpage
\begin{figure}
\centering
\mbox{\epsfig{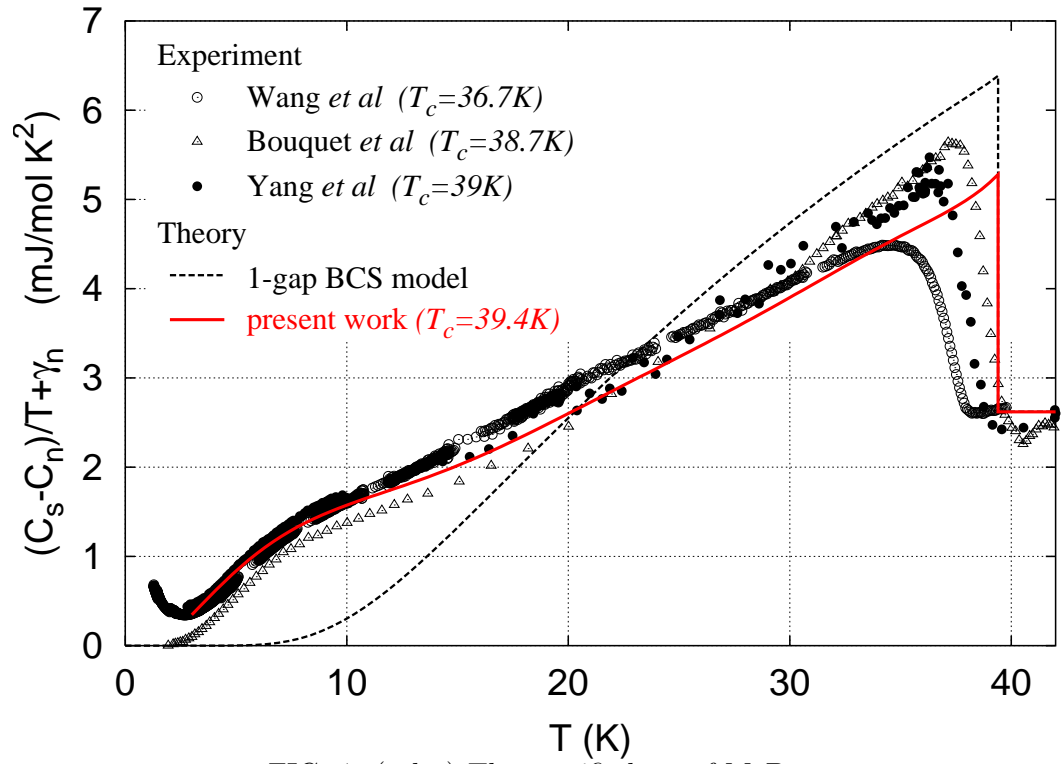}}
\caption{(color) The specific heat of MgB$_2$.}
\end{figure}

\end{document}